# Stacking transition in bilayer graphene caused by thermally activated rotation

Mengjian Zhu[1,†], Davit Ghazaryan[1,†], Seok-Kyun Son[2,†], Colin R. Woods[1], Abhishek Misra[1,2], Lin He[3], Takashi Taniguchi[4], Kenji Watanabe[4], Kostya S. Novoselov[1,2], Yang Cao[1,2,‡], Artem Mishchenko[1,2,‡]

[1]*School of Physics and Astronomy, University of Manchester, Oxford Road, Manchester, M13 9PL, UK*

[2]*National Graphene Institute, University of Manchester, Booth St. E, Manchester, M13 9PL, UK*

[3]*Department of Physics, Beijing Normal University, Beijing, 100875, China*

[4]*National Institute for Materials Science, 1-1 Namiki, Tsukuba, 305-0044, Japan*

[†]These authors contributed equally

[‡]Corresponding authors, e-mail: yang.cao@manchester.ac.uk, artem.mishchenko@gmail.com

**Abstract:** Crystallographic alignment between two-dimensional crystals in van der Waals heterostructures brought a number of profound physical phenomena, including observation of Hofstadter butterfly and topological currents, and promising novel applications, such as resonant tunnelling transistors. Here, by probing the electronic density of states in graphene using graphene-hexagonal boron nitride tunnelling transistors, we demonstrate a structural transition of bilayer graphene from incommensurate twisted stacking state into a commensurate AB stacking due to a macroscopic graphene self-rotation. This structural transition is accompanied by a topological transition in the reciprocal space and by pseudospin texturing. The stacking transition is driven by van der Waals interaction energy of the two graphene layers and is thermally activated by unpinning the microscopic chemical adsorbents which are then removed by the self-cleaning of graphene.

1. Introduction

A recent progress in two-dimensional (2D) materials and van der Waals (vdW) heterostructures has enabled the creation of artificial materials with tailorable properties for a range of different applications [1]. Although desirable in many cases, it is currently hard to achieve a perfect crystallographic alignment of individual layers in vdW heterostructures during stacking procedure [2]. In contrast, within parent bulk crystals the constituent layers are in a commensurate registry with each other because this optimises the interlayer vdW energy of the crystal. For instance, in mechanically exfoliated bilayer graphene the individual graphene layers are in a commensurate AB (Bernal) stacking, which minimises the vdW energy. This enables the interlayer electron hopping which dramatically modifies graphene band structure [3].

When bilayer graphene is prepared by a micromechanical transfer of two graphene layers on top of each other, there exists another degree of freedom – a twist angle between the graphene crystallographic directions. This rotational misalignment provides a tuning knob for the electronic spectrum of graphene bilayer. Twisting the two graphene crystals with respect to each other introduces a momentum mismatch between the two Dirac cones, which hinders the interlayer electronic coherence and, to some extent, preserves the linear energy dispersion [4].

When one monolayer graphene (MGr) is strongly misaligned with another MGr, the friction between two graphene layers can vanish, leading to a superlubric motion of the top layer [5-7]. Eventually, the vdW interaction energy will drive the system to the most energetically favourable commensurate state, the AB stacking bilayer graphene (BGr) [8, 9]. Similar phenomena (though driven by maximization of the overlapping area) indeed have been observed at the micron scale for self-retraction of three-dimensional graphite mesa structures [9, 10], and, more recently, for thermally induced self-alignment in MGr-hBN heterostructures [11, 12]. The reason why thermal activation is often necessary for macroscopic self-alignment is the presence of chemical adsorbents between the atomic layers, acting as nanoscale 'glue' and preventing the superlubric motion [13]. Such pinning may stabilise the crystals in an unstable configuration, such as twisted bilayer graphene (tBGr).

In the following, we demonstrate that, despite the strong pinning effect due to chemical contaminations between two graphene layers, tBGr can rotate towards stable BGr state after a long-time thermal annealing. In order to probe the structural transition from tBGr to BGr precisely, we employ the tunnelling spectroscopy technique using graphene-hBN-graphene tunnelling transistors, which is known for its high sensitivity to the density of states (DoS) of graphene [14].

## 2. Results and discussion

We fabricated graphene-based tunnelling transistors using a sequence of mechanical exfoliation and dry transfer procedures, similar to those described in previous works [14, 15]. Firstly, we used the standard micromechanical cleavage technique to prepare relatively thick (20–30 nm) hBN crystals on top of oxidised (290 nm $SiO_x$) silicon wafer which acted as a back gate. The hBN crystals served as a high-quality atomically flat substrate. We then transferred a first MGr flake on top of the selected hBN crystal using dry transfer procedure. A 3-layer hBN tunnel barrier was then identified and transferred on top of the first MGr. Finally, we completed the stack with the second MGr, using the same transfer procedure. After standard electron beam lithography and metallization (5 nm Ti / 50 nm Au), the structure was annealed at 200ºC in $Ar/H_2$ gas. The device schematics and micrograph are shown in figure 1.

In our tunnelling devices, the tunnelling current $I$ was measured as a function of applied bias voltage $V_b$ between the top and bottom graphene electrodes and the back gate voltage $V_g$ applied between highly doped silicon gate electrode and bottom graphene. In order to measure the differential tunnelling conductance $dI/dV$, we mixed small low-frequency ac voltage to dc bias $V_b$ and measured the ac current with a lock-in amplifier. Typical $I$-$V$ and $dI/dV$ characteristics for different $V_g$ measured at a base temperature of 1.6 K are shown in figure 1c.

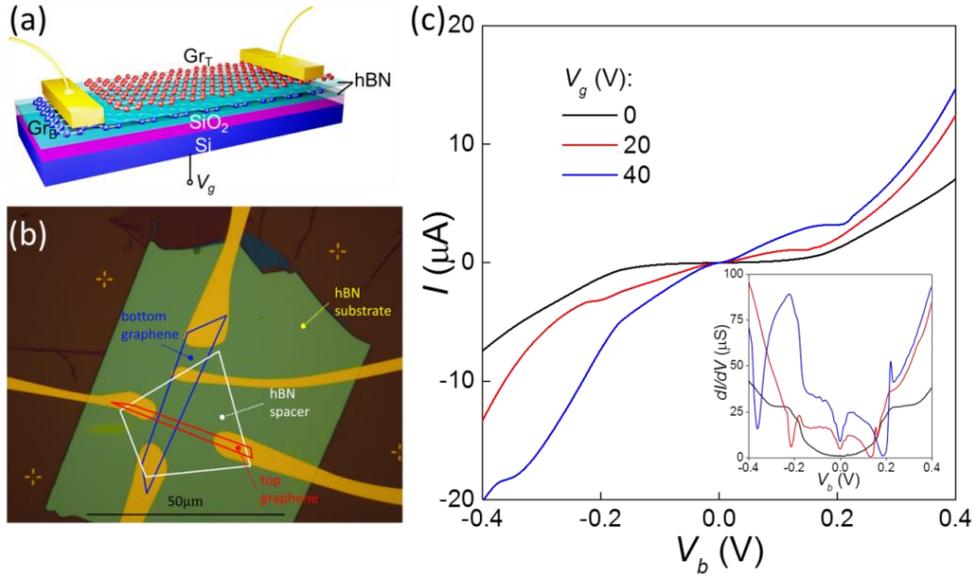

**Figure 1.** MGr-hBN-MGr tunnelling transistor. (a) Schematics and (b) optical micrograph of our device. A 3-layer hBN tunnel barrier (white outline) separates two MGr sheets. The overlap between top layer graphene (red outline) and bottom layer graphene (blue outline) corresponds to an active tunnelling area $A \approx 10\ \mu m^2$. (c) Tunnelling current $I$ as a function of $V_b$ at $T = 1.6$ K for different gate voltages. Inset shows the differential tunnelling conductance $dI/dV$.

The energy band diagram of our tunnelling device and the schematics of electrical measurements are shown in figure 2a. It is more informative to plot a contour map of $dI/dV$ as a function of $V_b$ and $V_g$, as shown in figure 2b. Here, the X-shaped white lines represent low tunnelling conductance, which we attribute to the passing of the chemical potential $\mu_i$ through the charge neutrality point (CNP) of each graphene layer. The $dI/dV$ is suppressed in this region due to the vanishing DoS.

We analysed the electrostatic properties of our tunnel transistors using a parallel-plate capacitor model [14]. In addition, we took into account that the electric field generated by the charge on the gate electrode is only partially screened by the bottom graphene layer, due to the low DoS in graphene. Based on this model, we have derived the following pair of simultaneous equations:

$$\begin{cases} eV_b = \mu_B - \mu_T - \Delta\varphi \\ eV_g = \mu_B - e^2 D(n_B + n_T)/\varepsilon\varepsilon_0 \end{cases} \quad (1)$$

Where $e$ is the elementary charge and $\mu_i$ and $n_i$ are the graphene chemical potentials and carrier densities respectively. In the case of MGr $\mu_i = \pm \hbar v_F \sqrt{\pi|n_i|}$. $D$ is the total thickness of $SiO_x$ and hBN substrate, $d$ is the thickness of hBN tunnel barrier, $\varepsilon_0$ is the vacuum permittivity and $\varepsilon$ is the medium permittivity (3.2 for hBN and 3.9 for $SiO_x$). Additionally, the band offset between the top and bottom graphene layers is obtained using $\Delta\varphi = e^2 d n_T/\varepsilon\varepsilon_0$.

We then fitted the experimental data with the described model and determined the conditions for the intersection of the chemical potentials with the CNP for each graphene layer. The fits are shown as dashed blue and red lines for top and bottom graphene layers respectively (figure 2b). Note that Fermi velocity $v_F = 1.05 \cdot 10^6$ m/s extracted from our electrostatic model is in a good agreement with the literature [16].

Apart from the CNP-related low DoS features, there are other regions with suppressed electronic DoS, and we can clearly distinguish them from a darker background as sharp vertical (independent of $V_g$) lines at $V_b \approx 12, 20, 170$ and 200 mV, figure 2b. We attribute those features to phonon-assisted tunnelling processes where conductance increases as a series of steps every time $V_b$ is large enough to excite a phonon $eV_b = \hbar\omega_{ph}$ [17, 18].

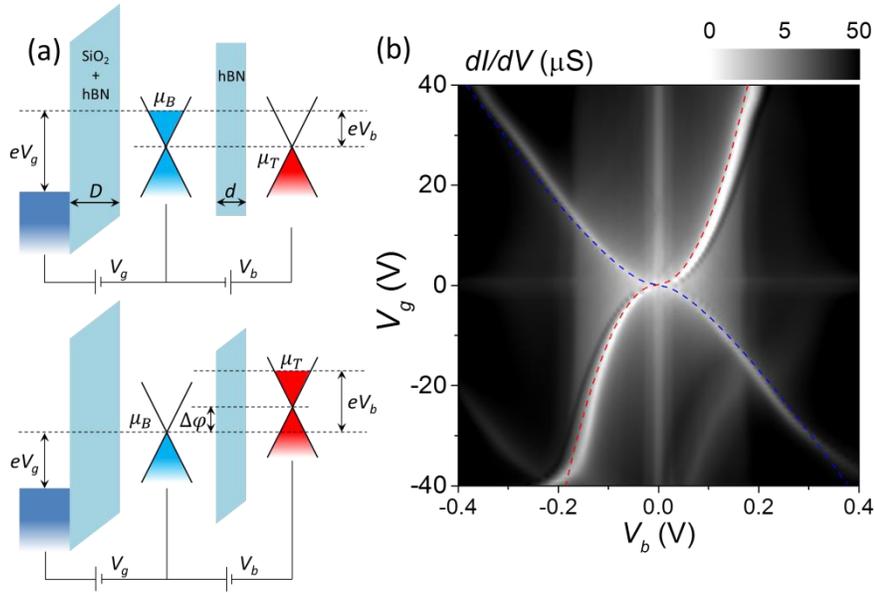

**Figure 2.** Differential tunnelling conductance of MGr-hBN-MGr tunnelling transistor. (a) Schematics of energy band diagrams of the MGr tunnelling transistor when the chemical potential in one graphene layer passing through its CNP. (b) Measured *dI/dV* map as a function of $V_b$ and $V_g$. Red and blue dashed lines represent the electrostatic calculations for zero chemical potential in the top and bottom graphene, respectively. Their respective band diagrams are presented on top and bottom schematics of panel (a).

Having characterised our tunnelling transistor we then transferred another, smaller, MGr flake on top of the upper graphene electrode using the same dry transfer method. During the transfer, we intended to align the graphene flake within 5° with the top graphene layer to obtain tBGr as the top electrode.

Rotational misalignment between two MGr crystals leads to a moiré pattern, as depicted in figure 3a. The superlattice period $\lambda$ is determined by the twist angle $\theta$: $\lambda = a_0/2\sin(\theta/2)$, $a_0$=0.246 nm is the graphene lattice constant [19]. In addition to the structural modification, such rotation causes a shift between the Dirac points in the reciprocal space by $|\Delta \mathbf{K}| = 2|\mathbf{K}|\sin(\theta/2)$, where $\mathbf{K}$ is the reciprocal lattice vector with $|\mathbf{K}| = 4\pi/3a_0$, leading to a significantly different spectrum as compared to both monolayer and bilayer graphene, as demonstrated in figure 3b. If there is a finite interlayer coupling between the two MGr layers, two saddle points will apear at the intersection of the two cones resulting in two van Hove singularities (VHS) in the DoS (figure 3c,d) at energies of $E_{VHS}^{\pm} = \pm(\hbar v_F |\Delta \mathbf{K}|/2 - \tau_\theta)$, where $v_F$ the Fermi velocity of MGr and $\tau_\theta$ the interlayer hopping parameter [4, 20-23].

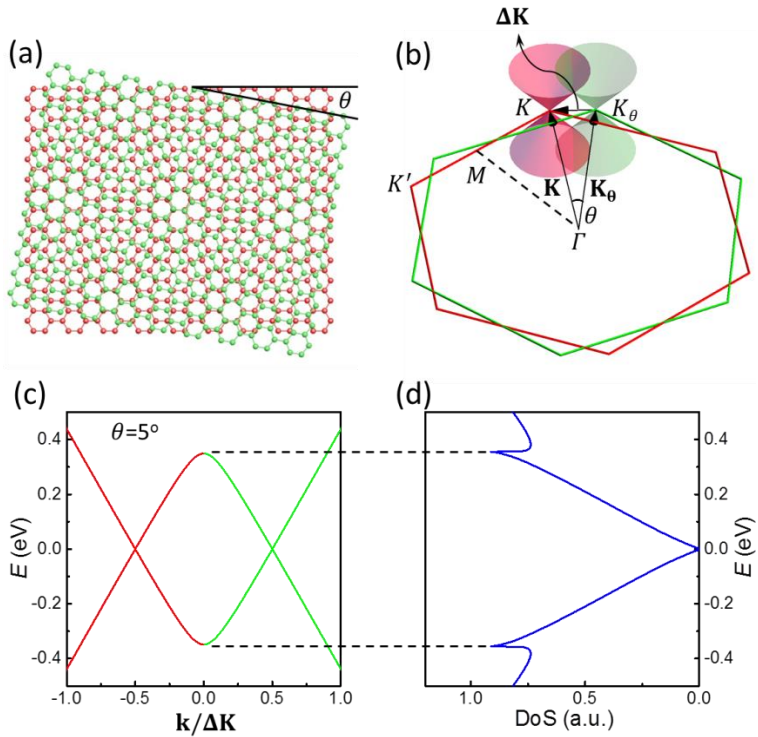

**Figure 3.** Twisted bilayer graphene. (a) Schematic structural model of two misoriented honeycomb lattices with a twist angle $\theta$. (b) Schematic Dirac cones of the two graphene layers in the reciprocal space at $K$ points. High-symmetric points are also shown here. (c) Electronic band structure of tBGr with wavevector **k** varying from $-\Delta \mathbf{K}$ to $\Delta \mathbf{K}$ along the line joining the two Dirac cones. $\theta = 5°$ and $\tau_\theta = 110$ meV are used in the calculations following [22, 23]. (d) Calculated corresponding low-energy DoS in twisted bilayer graphene. Two van Hove singularities are indicated by the dashed lines.

After transfer, we carried out the same tunnelling spectroscopy measurements, see figure 4e. Noticeably, CNP of the top tBGr graphene electrode shifted from 0 V to 10 V after transfer. This is a strong evidence of the presence of adsorbents and chemical residues left after device fabrication in-between top graphene electrode and recently added graphene flake. This 'glue' prevents the superlubric motion and rotation of the top MGr flake immediately after the transfer and stabilises twisted bilayer configuration.

Crystallographic orientations of the flakes and corresponding tunnelling spectroscopy maps are shown in figure 4b and e, where for consistency we used $\Delta V_g$ instead of $V_g$ to characterise the carrier density in graphene electrodes. The *dI/dV* map of tBGr-MGr (figure 4e) compares well with MGr-MGr tunnelling device (figure 4d). That is, the X-shaped feature is also present in figure 4e indicating a low energy linear dispersion in tBGr, which is qualitatively similar to the MGr case. Quantitatively, there is a significant difference between of tBGr and MGr tunnelling devices when the chemical potential of bottom MGr sits at zero $\mu_B = 0$. We attribute the shrinked X-shaped feature in tBGr tunnelling device to the increased DoS in tBGr electrode. At low energy, there are two Dirac cones at each corner of the Brillouin zone of tBGr instead of single Dirac cone for the MGr case. In this simple assumption we ignored the Fermi velocity renormalization in tBGr due to a relatively large twist angle ($\theta \sim 5^o$) and the unknown interlayer coupling strength [19, 24]. The calculation agrees well with our experimental data for tBGr tunelling devcie (dashed green line in figure 4e) after we modified the electrostatic model by adding an equation to describe the second top MGr layer ($T_2$) electrically shorted with the first top MGr layer ($T_1$):

$$\begin{cases} eV_b = \mu_B - \mu_{T1} - \Delta\varphi_1 \\ eV_g = \mu_B - e^2 D(n_B + n_{T1})/\varepsilon\varepsilon_0 \\ \mu_{T1} = \mu_{T2} + \Delta\varphi_2 \end{cases} \quad (2)$$

Here $\Delta\varphi_1 = e^2 d(n_{T1} + n_{T2})/\varepsilon\varepsilon_0$ is the band offset between the bottom, MGr, and the top, tBGr, electrodes separated by hBN and $\Delta\varphi_2 = e^2 d' n_{T2}/\varepsilon_0$ is the band offset between $T_1$ and $T_2$ where $d'=0.35$nm is the vdW gap between the two graphene layers. We did not obeserve any signatures of VHS in tBGr tunnelling device. We attribute this to a large twist angle ($\approx 5°$) which shifts VHS to above 0.4 eV, higher than the chemical potential (~ 0.2 eV) in tBGr we can reach by applying gate and bias voltages.

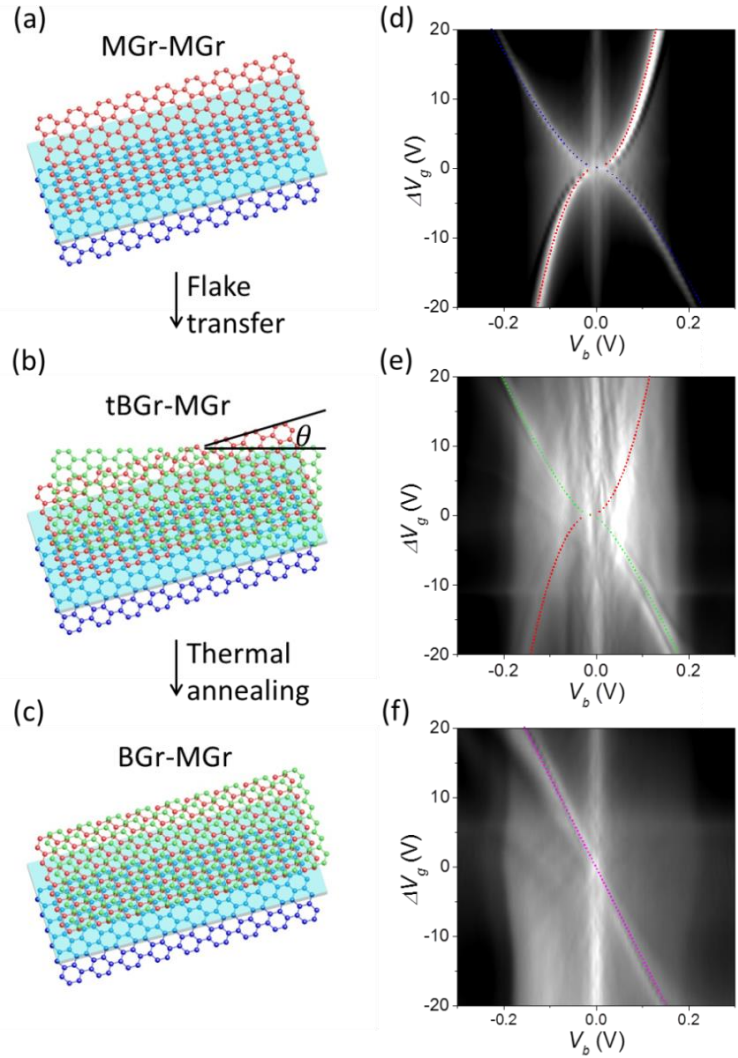

**Figure 4.** The evolution from MGr to tBGr and to BGr. (a)-(c) Schematics of different graphene tunnelling transistors (only graphene-hBN-graphene heterostructures are shown). (d)-(f) measured $dI/dV$ maps as a function of $V_b$ and $V_g$ for MGr-MGr, tBGr-MGr and BGr-MGr tunnelling devices. Dashed lines are the electrostatics calculations for zero chemical potentials.

Finally, we annealed the device at 200ºC for 12 hours under Ar/H$_2$ and then carried out the same tunnelling spectroscopy measurements. The CNP of the top graphene layer shifts from 10 V to 3 V, suggesting that chemical adsorbents were partially removed by thermal annealing. Measured *dI/dV* map (figure 4f) clearly shows two differences in comparison with tBGr case (figure 4e). First, the zero DoS line for the bottom graphene becomes linear rather than square root shape, in contrast to both MGr-MGr and tBGr-MGr cases. Secondly, the zero DoS line for the top graphene disappears. These two differences can be explained by the transition of the top graphene electrode from tBGr to BGr after thermal annealing. To quantify these findings we fit the linear zero DoS line (see figure 4f) with modified electrostatic model considering AB stacking bilayer graphene as the top electrode [25]:

$$\begin{cases} eV_b = \mu_B - \mu_T - e^2 d n_T/\varepsilon\varepsilon_0 \\ eV_g = \mu_B - e^2 D(n_B + n_T)/\varepsilon\varepsilon_0, \\ \Delta = e^2 d'(n_B + n_T'')/\varepsilon\varepsilon_0 \end{cases} \qquad (3)$$

where the relation between carrier density and chemical potential is given by $n_T(\mu_T, \Delta) = \frac{m}{\pi} Re\sqrt{(2\mu_T)^2 - \Delta^2}$, here $m$ is the effective carrier mass, $\Delta$ the tunable bandgap, $n_T'' = n_T - n_T'$ the carrier density on one particular layer in AB stacking bilayer.

To further confirm observed tBGr-to-BGr transition we performed another annealing experiment using Raman spectroscopy (Renishaw Raman spectrometer, excitation line of 514.5 nm, and incident power of 2 mW) as an independent tool to identify a twist angle between the top and bottom MGr flakes. The full width at half maximum (FWHM) of a 2D peak (~2690 cm$^{-1}$) of tBGr is sensitive to the twist angle ($\theta$), increasing from ~20 cm$^{-1}$ for $\theta > 15º$ up to ~50 cm$^{-1}$ for $\theta = 0º$ (BGr) [26]. We fabricated tBGr sample on hBN substrate (figure 5a) and measured its Raman maps of FWHM of 2D peak before and after annealing (figure. 5b-c). Maps clearly show a difference in $\theta$ after annealing: the average FWHM was 45.3±0.9 cm$^{-1}$, which corresponds to ~4-5º twist, while after annealing FWHM increased to 50.0±0.9 cm$^{-1}$ – a value typical for BGr on hBN [26]. Interestingly, a small part of a top MGr has rotated in the opposite direction after annealing (FWHM 33.8±1.7 cm$^{-1}$ corresponding to ~9-10º twist), probably to compensate for a strain due to the presence of two large bubbles in that region of the tBGr. As it was demonstrated recently, the average strain in graphene enveloping bubbles is ~1% at its top, independent of the size and shape of the bubble and of the properties of trapped material [27].

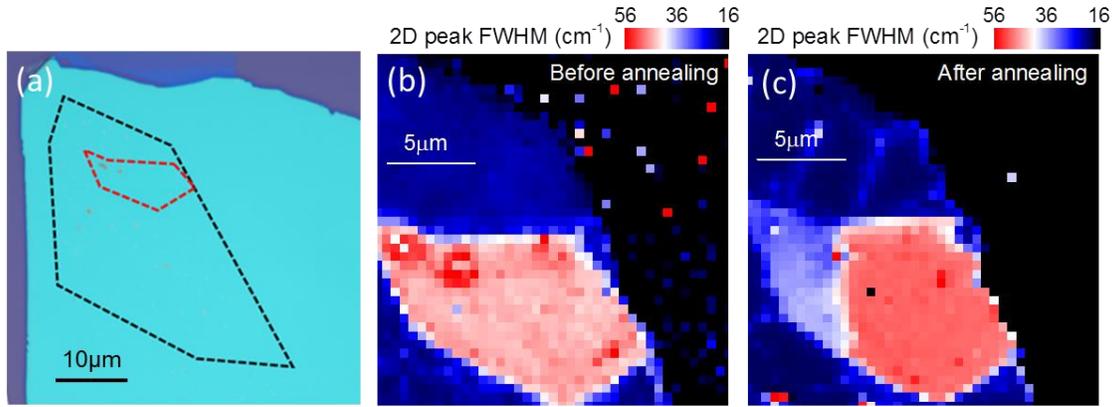

**Figure 5.** (a) Optical micrograph of tBGr on hBN heterostructure. The red and black dashed lines indicate the edges of top and bottom MGr flakes respectively. (b) and (c) Raman maps of FWHM of 2D peak before (b) and after (c) annealing.

Van der Waals energy between graphene layers strongly depends on the misalignment – the minimum energy corresponds to AB stacking at every 60º intevals [28, 29]. The energy difference between AB stacking and a fully misaligned ($\theta$ = 30º) conformation is ~0.13 meV/atom [30]. This vdW energy difference is sufficient to drive a transition from an incommensurate stacking state to a commensurate AB stacking since they are separated by only a small barrier <0.01 meV/atom [30]. The reason the transition doesn't usually happen before thermal annealing is most likely pinning effects due to the chemical adsorbents between the graphene layers [31]. Indeed, it was recently demonstrated that the dissipative energy, related to irreversible removal of contaminants from the graphite-graphite interface, decreases rapidly with increasing temperature [32]. During the long-time thermal annealing, those contaminants become mobile and then are segregated to localised pockets by the self-cleaning driven by vdW force between two graphene layers [11, 33]. Remarkably, the observed tBGr to BGr transition implies that the top 5 μm size MGr has rotated by at least 500 nm [11], which can find applications in nanoelectromechanical systems.

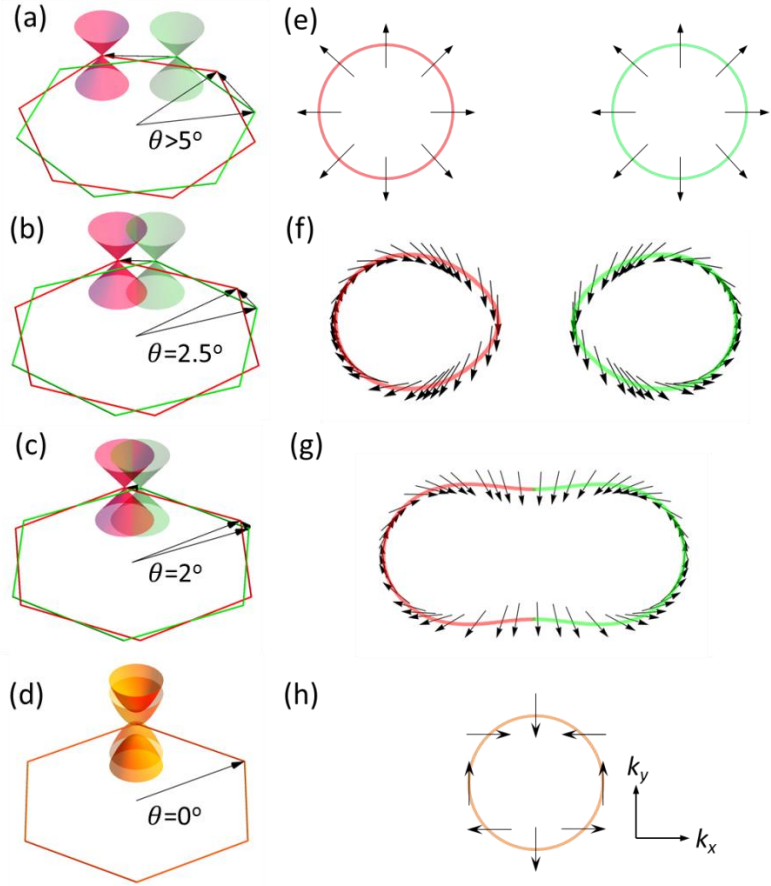

**Figure 6.** Structural transition in bilayer graphene from tBGr to BGr. **(a)-(c)** Schematics of the evolution of Brillouin zone from the large-$\theta$ tBGr to small-$\theta$ tBGr to BGr **(d)**. Low energy dispersions (Dirac cones) are shown only for one corner of the Brillouin zone for top (red) and bottom (green) graphene layers, Twist angles $\theta$ are exaggerated. **(e)-(h)** Low-energy ($E = 60$ meV) Fermi surfaces and pseudospin texture *vs.* twist angle $\theta$ for $K$ valley, calculated as described in [34]. The interlayer hopping $\tau_\theta = 110$ meV was used. Enclosed curves indicate the Fermi surface topology and black arrows represent the pseudospin angle.

The observed incommensurate-commensurate transition between tBGR and BGr was predicted to occur under certain conditions, such as strain in one of the layers [35, 36]. In our work we present an alternative route to achieve this transition by graphene-to-graphene self-rotation, see figure 6. The tBGr behaves as two decoupled MGr when the twist angle is large enough and the interlayer interaction can be neglected (figure 6a). During annealing the top MGr rotates, driven by the gradient in vdW energy. As a result, $\theta$ decreases and two saddle points apear at the intersection of the two Dirac cones resulting in two VHS in DoS (figure 6b, c). At low enough angles ($\theta < 1.5°$, where the moiré period becomes > 10 nm) the picture rapidly gets complicated due to the reentrant appearance of flat moiré bands with vanishing Fermi velocity at several commensurable twist angle values [37, 38]. Finally, graphene reaches the BGr structure, forming a parabolic low-energy dispersion (figure 6d).

This structural transition is also accompanied by the topological transition in the reciprocal space - the twist dependent Lifshitz transition, where the rotation modifies the Fermi surface topology of the system: two independent Fermi surfaces merge into one after the rotation, figure 6 e-h. Analogous topological phase transitions were observed in tBGr and BGr in high magnetic fields [39, 40], while in our case it is driven by a mechanical rotation of the two graphene crystals. In addition to topology, a pseudospin texture is also affected by the structural transition. The electron wave function in graphene consists of two components, which represent the probability of finding an electron on the two sublattices of the honeycomb lattice. The pseudospin is the phase shift between those components, which, for monolayer graphene is locked to the direction of the electron's motion.

The tBGr with large twist angle ($\theta > 5°$) can be treated as two decoupled MGr layers with the same pseudospin texture, as shown in figure 6e. For smaller-$\theta$ tBGr, due to the interlayer interaction, the pseudospin vectors rotate to lower the symmetry of tBGr, which merges the two Fermi surfaces above the VHS (figure 6f, g) [34, 41]. Finally, the tBGr system reaches its commensurate state when $\theta = 0°$ and two MGr aligns with each other in AB stacking, where the Fermi surfaces completely merge to single one and the pseudospin texture assumes the BGr case (figure 6h). This evolution of pseudospin texture also implies a continuous Berry phase transition from $\pi$ for large-$\theta$ tBGr) to $2\pi$ (BGr) through small-$\theta$ tBGr state [39, 40]. Our results (as shown in figures 4 and 6) provide the first experimental evidence of a thermally activated structural transition in bilayer graphene driven by the self-rotation, which paves the way to nanomechanical band structure engineering.

## 3. Conclusions

To summarise, we demonstrated the structural transition from twisted bilayer graphene to AB stacking bilayer graphene using the highly sensitive technique of tunnelling spectroscopy. We also showed that thermal annealing unpins the graphene allowing its macroscopic self-rotation driven by vdW energy. We believe this technique will be useful to study similar structural transitions in nanoelectromechanical systems, clearing the way to mechanically driven band structure engineering.

**Acknowledgments**

This work was supported by the European Research Council, the EU Graphene Flagship Program, the Royal Society, the Air Force Office of Scientific Research, the Office of Naval Research and ERC Synergy Grant Hetero2D. Mengjian Zhu acknowledges the National University of Defence Technology (China) overseas PhD scholarship. Colin Woods acknowledges the support of the EPSRC (UK). Artem Mishchenko acknowledges the support of EPSRC (UK) Early Career Fellowship EP/N007131/1.